# A Formalizable Proof of the No-Supervenience Theorem: A Diagonal Limitation on the Viability of Physicalist Theories of Consciousness.

**Cathy Reason**

**London, United Kingdom**

*The no-supervenience theorem limits the capacity of physicalist theories to provide a comprehensive account of human consciousness. The proof of the theorem is difficult to formalize because it relies on both alethic and epistemic notions of possibility. This article outlines a formalizable proof using predicate modal logic in which the epistemic inferences are expressed in terms of an existing mathematical formalism, the inference device (Wolpert, 2008). The resulting proof shows definitely that any physicalist theory which describes a self-aware, intelligent system must be internally inconsistent.*

## Introduction

The no-supervenience theorem is a no-go theorem which establishes constraints on theories of consciousness by examining the general properties of sets of theories, without having to take into account the metaphysical qualities associated with individual members of those sets. The procedure for doing so was first outlined by Caplain (1995, 2000). Subsequent authors developed Caplain's ideas into an informal proof, showing that no physical system capable of humanlike reasoning could perform the Cartesian cogito (Reason 2016, 2019; Reason and Shah 2021). Similar, though inequivalent, arguments have since been advanced by Gao (2024), Kim (2024), and List (2023, 2024).

In this article, we shall illustrate the proof with a simple thought experiment, the Three World Problem, and then present a formalizable proof of the no-supervenience theorem. The formalizable proof exploits an existing mathematical formalism, the *inference device* (Wolpert 2008). Wolpert showed that all physical systems capable of observation, prediction and memory have a common mathematical structure. Any system with such a structure can be classed as an inference device. For our purposes, the exact details of this structure need not be considered; it is enough to define an inference device as a general mapping, which we shall call the *Wolpert mapping.* The concept of an inference device allows epistemic modal inferences to be codified formally without requiring an explicitly epistemic interpretation of modal logic. This allows the required inferences to be expressed formally in predicate modal logic with set theory, and avoids the problem of combining epistemic and alethic modal interpretations in a single proof.. It should be noted that the notion of a physical system we use is considerably simpler and more limited in scope than Wolpert's original formalism, but is based on the same basic concept, that inference can be represented in terms of mappings within physical systems.

**The Three Worlds Problem: A Brief Illustration of the Proof**

We shall now start with a simple thought experiment.

Let us define the proposition P:

P: "I am not currently in a dreamless sleep."

We shall call this the *Cartesian proposition.* Note that alternative versions of P will do equally well, including those which make no reference to the concept of "I". For simplicity we shall use the definition above without loss of generality.

Now let us define the concept of physicalism. A physicalist theory is one in which every mental process implies the existence of some physical process. We shall define a physical process to be one which exists independently of any given observer. (In the formalizable proof below, we shall employ a different definition of objectivity, but this will do for the moment.) We shall define the proposition Q as follows:

Q: "The physical process necessary for P to be true, both exists and operates correctly."

For physicalism to be true, it must be the case that P implies Q. We can therefore define three epistemically possible cases, or *worlds*:

W1: True(P) False(Q)
W2: True(P) True(Q)
W3: False(P) False(Q)

Clearly, worlds W2 and W3 are consistent with physicalism, but world W1 is not. Now the problem of asserting P can be expressed as the problem of reducing these three epistemically possible worlds to two, such that W1 and W2 are epistemically possible, but W3 is not (since we cannot assert P if there is an epistemically possible world in which P is false). If physicalism is true, then there must be some physical process which is responsible for this reduction, which we shall call X. Now clearly, every physical process is potentially fallible, so the existence and correctness of X cannot be decided a priori. We can therefore express the three worlds after application of this process as follows:

W1 True(P) False(Q)
W2 True(P) True(Q) Correct(X)
W3 False(P) False(Q) Incorrect(X).

Clearly we still have three epistemically possible worlds, because it is epistemically possible that X does not work correctly. So to reduce the three worlds to two, we need either another application of the process X, or some new process. We shall call this new process (or new application of an existing process) X1. Since X1 is a physical process and therefore fallible, we end up with the resulting set of worlds:

W1: True(P) False(Q)
W2: True(P) False(Q) Correct(X) Correct(X1)
W3: False(P) False)Q) Incorrect(X) Incorrect(X1)

It is apparent that we have still not been able to reduce the three worlds to two. Furthermore, it should now be obvious that we will *never* be able to reduce the three worlds to two, no matter how many times we apply the process X. or how many new processes Xn we apply to the set of three worlds. *And this conclusion is completely independent of any assumptions about how the physical processes operate, or what metaphysical properties they may have.* There is simply no way, using any set of physical processes, however large, that the set of three epistemically possible worlds can be reduced to two, and therefore no way, if physicalism is epistemically possible, that P can ever be assumed to be true, since there will always be an epistemically possible world in which P is not true. It is not possible, therefore, for any physicalist theory to describe any sort of system or being which is capable of any sort of Cartesian self-awareness, no matter how that is defined or interpreted.

We shall now express this argument as a formalizable derivation using modal logic.

## A Note on the Use of Modal Logic

The proof uses modal logic, and in particular relies on a form of the following equivalence:

$|\!-p \leftrightarrow \sim\Diamond\sim p$

(Actually the form we shall use will be slightly expanded, as explained below.) Afficionados of modal logic will doubtless recognize that this equivalence -- which asserts that no proposition can be necessarily true in a given language unless it is a theorem of that language -- is not generally true. However it *is* true in any axiomatization of modal logic where *only* theorems of modal logic are allowed as axioms -- or what we shall henceforth call a *restricted axiomatization*. For example, a model M comprising two possible worlds:

w1: P and Q
w2: P and ~Q

does not support the inference:

$(P \supset Q) \supset (\Diamond\sim Q \supset \Diamond\sim P)$

But this is only true because the model implicitly assumes $\sim\Diamond\sim P$ as an axiom, and since this axiom is not a theorem of predicate modal logic, it will not be necessarily true in any restricted axiomatization of predicate modal logic. One can express this by saying that if the axiom or axioms which define a model are not necessary, then any implication contingent on those axioms is not generally necessary either, or alternatively, that if, for some axiom P which defines a model:

$\sim P \supset \Diamond Q$

then:

$\Diamond\sim P \supset \Diamond Q$

It therefore necessarily follows that the equivalence:

$|\text{-}p \leftrightarrow \sim\Diamond\sim p$

can only be invalidated if additional axioms or postulates are added to the restricted axiomatization. The effect of this restriction is to limit the application of modal reasoning to a priori inferences. The restricted axiomatization is also effectively *monotonic*, which is to say, any valid argument in the restricted axiomatization cannot be invalidated by any subsequent argument. Any propositions which are not theorems of the modal logic will therefore have to be inferred empirically, and for this we use the mechanism of the Wolpert inference device. Any full, unrestricted axiomatization is clearly not monotonic since it allows for the empirical inference of models which are not theorems. It is on this key point which the proof ultimately depends. In the following, proof, we shall assume that any system can rely on its ability to reason using the restricted axiomatization of predicate modal logic -- which is to say,. the restricted set of axioms and all theorems deducible from them.

This obviously prohibits the use of axioms to represent so-called "self-evident" or empirical propositions. Any such propositions outside of the restricted set must be inferred using the mechanism of the Wolpert inference device. In order to do this, we must expand the above equivalence to include necessities inferred by Wolpert devices, so as to allow for the possibility of "a posteriori" necessities (Kripke 1986):

$(|\text{-}p \vee W(\sim\Diamond\sim p)) \leftrightarrow \sim\Diamond\sim p$

The "W" notation will be explained further below[1].

## The Proof in Formalizable Form

We can express the proposition that the Cartesian Cogito is generally impossible in any physicalist theory, in the terms of the following theorem and corollary:

*Theorem:* Any physical inference device, capable of reasoning consistently using predicate modal logic with set theory, can infer that none of its inferences are epistemically reliable.

*Corollary:* No physical inference device, capable of reasoning consistently using predicate modal logic with set theory, can ever infer that it is conscious, irrespective of how consciousness is defined.

*Proof:* We first assume some restricted axiomatization of first-order modal predicate logic.[2] Next we define a set Z of what may be called "facts about the world", and denote by p(Z) any partition of this set which contains only facts which are true at some time t. The subset p(Z) can be regarded as the set of facts in Z which are true whenever some proposition p is true. We also define a physical system as some set of states, together with a set of dynamical rules linking those states. We shall define M(t) to be the state of system M at time t, where t can be thought of as simply an arbitrary parameter. For convenience, the parameter t will subsequently be dropped from the notation. We shall assume that M is capable of reasoning using predicate modal logic.

[1] Strictly, we will need to add a further condition for this equivalence to hold -- that the Wolpert mapping is a process of Type P. This will be explained below.

We next define an *observer*, which we represent as a simplified version of a Wolpert inference device. An observer is a physical system O which, when working correctly, evolves to a special state W(p) (or the *Wolpert* state for p) such that O evolves to the state W(p) only if p is true. Whenever O evolves to W(p) and p is true, we shall say that W(p) has the value TRUE. In future we shall drop the notation p(Z) and simply use p to denote the subset of Z which is true whenever the proposition p is true. We shall define a *physical process* to refer to any subset or subcategory of a physical system. We can define a Wolpert state as *accurate* and *reliable* for some proposition p using *modal logic* if:

$$\sim\Diamond\sim(W(p) \supset p)$$

This simply says that, if the device O is operating correctly, then O cannot evolve to the state W(p) unless p is true. Informally, we can say that W(p) is the state to which some device O must evolve if O has asked the question "Is p true?" W(p) is equivalent to the answer YES to this question, and hence W(p) represents O's knowledge of the facts denoted by p.

We next need to define the notion of a *real* process, as opposed to a merely physical one. We define the set of *real* processes to be the set of all processes $\Phi$ such that, for two observers $O_A$ and $O_B$, the Wolpert states $W_A(\Phi \neq \emptyset)$ and $W_B(\Phi \neq \emptyset)$ both have the value TRUE, and $A \neq B$. More generally, we can define a *real property* to be any subset x of Z such that $W_A(x)$ and $W_B(x)$ both have the value TRUE and, $A \neq B$. This more general definition we shall refer to as the *realism postulate*[3]. We shall also define the set of *subjective* processes to be the set of all processes $\Psi$ for which the realism postulate does not hold. This somewhat complicated bit of notation simply expresses rigorously the notion that any *real* process should be an *objective* process, which is to say, it should, in principle, be observable by multiple witnesses. (Alternative definitions of objectivity can be used without loss of generality.)

We now define a *physicalist theory* as any theory which stipulates that every subjective process implies the reality of some real process, which is to say:

$$\Psi \supset \Phi$$

---

[2] A fully formal proof would also require some axiomatization of set theory, since sets play a role in the proof. However, since set theory can itself be expressed in first-order predicate logic, we shall make no further mention of this and include the axioms of set theory within the restricted axiomatization.

[3] There is no need to add this postulate, or the subsequent physicalism postulate, to the restricted axiomatization. They should instead be regarded as the antecedents of material conditionals.

We shall refer to this as the *physicalism postulate*. We shall describe as *physicalist* any theory which satisfies this axiom.

Now let X be some real process. We shall say that X is a process of type P if it performs a mapping (the *Wolpert mapping*) such that:

$O \rightarrow W(p)$ if and only if p is true.

We shall say that X is a process of type R if it performs a mapping (the *anti-Wolpert* mapping) such that:

$O \rightarrow W(p)$ if and only if p is untrue.

Let S(X) be any set of real processes sufficient to enable a Wolpert device to establish some proposition p, subject to the condition that every X is a process of type P -- which is to say, not a process of type R.

Now define the proposition "X is not a process of type R", or:

$X \neq R$[4]

Now it is clear from the postulates of realism and physicalism, that if X is real and X = P , then P must also be real and so X = P must be a real property. Since X = P implies $X \neq R$, this means $X \neq R$ is also a real property[5]. But no single observer can prove *a priori* if a property is observable by others since

$(O_A \neq \emptyset) \supset (O_B \neq \emptyset)$

where $A \neq B$, is not provable in the restricted axiomatization. This can be expressed in modal logic as:

$\Diamond\sim((O_A \neq \emptyset) \supset (O_B \neq \emptyset))$

Hence the inference:

$W(X \neq R) \supset (X \neq R)$

is not generally valid. From this it follows that:

$W(p) \supset p$

is not generally valid either. We now have two possibilities to consider -- the case where p is alethically necessary, and the case where p is not alethically necessary. We will sart with the case in which p is not alethically necessary According to the restricted axiomatization of predicate modal logic, we can derive::

---

4 Note the condition $X \neq R$ is automatically satisfied if $X = \emptyset$ -- in other words, if X does not exist.

5 One can assume for simplicity that X= P and $X \neq R$ are equivalent, but it is not actually necessary.

$\lozenge\sim(W(p) \supset p)$[6]

We shall refer to this fact as the *principle of empiricism*. In plain English, this principle says that no single observer can infer *a priori*, from the restricted axioms of predicate modal logic, whether any given real process is a process of type P or of type R. We can express this using modal logic as:

$\lozenge(W(p) \;\&\; \sim p)$

Now, we shall stipulate that M can reason using first-order predicate modal logic. It follows from any restricted axiomatization of predicate modal logic that:

$W_M(\lozenge(W_M(p) \;\&\; \sim p)) = \text{TRUE}$

From which it follows that:

$W_M(\lozenge\sim(W_M(p) \supset p) = \text{TRUE}$

And so $W_M(p)$ clearly contradicts the definition of a reliable Wolpert state. So we can conclude that, given the restricted axiomatizations of predicate modal logic with the postulates of physicalism and realism, no real physical system which operates as a Wolpert inference device can accurately infer p.

To get around this, one would have to add additional axioms, such $X \neq R$ and $W_M(X \neq R) = \text{TRUE}$, to the restricted axiomatization. But by the physicalism postulate, any Wolpert state such as $W_M(X \neq R) = \text{TRUE}$ implies some real process $X_W$ to perform the necessary Wolpert mapping, and by the principle of empiricism, there is no *a priori* way that M can be sure that $X_W$ is a process of type P. Indeed, for any set of processes S(X), it will always be the case that:

$\lozenge(X = R)$ for all $X \in S(X)$

unless $(X \neq R)$ for all $X \in S(X)$ is already known. In which case there must be some process $X_{diagonal}$ which maps M to the Wolpert state:

$W_M(W_M(X \neq R) = \text{TRUE for all } X \in S(X))$

Now, if $X_{diagonal}$ is a member of S(X), then M can only be sure that $X_{diagonal}$ is accurate by assuming it axiomatically, since $X_{diagonal}$ is capable of generating the state:

$W_M(W_M(X_{diagonal} \neq R) = \text{TRUE})$

regardless of whether $X_{diagonal}$ performs a Wolpert mapping or an anti-Wolpert mapping. In formal terms this means that:

6 This equivalence applies only if no further relevant axioms are assumed. For example, Kripke (1986) has proposed an additional axiom of the form $W(q) \rightarrow \sim\lozenge\sim q$ for certain values of q (the case of so-called a posteriori necessities). However this axiom is not a theorem of predicate modal logic, and so, by the principle of empiricism, is not necessarily true unless yet further axioms are assumed.

$\Diamond(W_M(W_M( X_{diagonal} \neq R ) = TRUE) \neq TRUE)$

is provable in predicate modal logic, given the postulates of realism and physicalism.

We now need to consider the case where p is alethically necessary. In such a case, there obviously can be no process of type R by which a Wolpert device can evolve to the state W(p = TRUE) only when p is false, since p is never false. However, we note that there is no axiom in the restriced axiomatization such that:

$\sim\Diamond\sim p \supset W(\sim\Diamond\sim p)$

and consequently, there is no such axiom as

$\sim\Diamond R \supset W(\sim\Diamond R)$

either. Therefore, $\sim\Diamond R$ cannot be assumed by M unless it is inferred by a Wolpert inference. It follows that the conclusion

$\Diamond(W_M(W_M( X_{diagonal} \neq R ) = TRUE) \neq TRUE)$

is provable in predicate modal logic *even in those cases where p, and hence R, are alethically necessary.* Or in plain English, it is provably possible that whatever M "knows" about the reliability of its own knowledge is, simply, wrong. It necessarily follows that, if M is capable of reasoning in the restricted axiomatization, then M itself can prove that all of its knowledge could be wrong. One could only circumvent this by adding further axioms, such as unconditionally assuming that $X_{diagonal}$ must be reliable. But unconditionally assuming the reliability of $X_{diagonal}$ clearly violates the physicalism postulate. And if $X_{diagonal}$ is not a member of S(X), then this violates the definition of S(X). So we can say definitively that no system which satisfies the definition of physicalism, and is capable of reasoning using predicate modal logic, can establish any empirical proposition p. And since we can define some proposition q thus:

q: "The proposition p can be established as true with probability of at least k"

where k is a real number $0 \leq k \leq 1$, we can also prove in predicate modal logic that M cannot establish any empirical proposition p with any quantifiable degree of confidence, without making additional axiomatic assumptions which are provably incompatible with physicalism.

## Some General Remarks on the Theorem

The first point to note, is that the diagonal contradiction arises from the stipulation that real processes are objective. Although the expression W(p) = TRUE also induces a potentially infinite number of Wolpert states, this does not imply a contradiction, since the Wolpert state is a theoretical construct and any number of Wolpert states could in principle be generated by the same physical process. But because the physical processes are required to be objective, the principle of empiricism induces a non-terminating sequence of physical processes, and this is what generates the contradiction.

Secondly, we note the following corollary of the theorem proved above. Should there exist some subset $\omega(Z)$, such that $W_M(\omega)$ = TRUE can accurately be determined by M even when $\omega$ is not alethically necessary, then M is performing some function which is inconsistent with any physicalist theory. Reason and Shah (2021) referred to such a function as an *omega function*. One example of such a function which can be performed by conscious human beings is the Cartesian cogito -- in fact, mathematically, $\omega$ is equivalent to the proposition P defined in the three world problem above. There are numerous different interpretations of the cogito, but for our purposes, we can describe a system as *omega-capable* if it can perform just one of them, and it does not matter which one. For simplicity, we shall say that a conscious human being is omega-capable if they can determine accurately that they are not in a dreamless sleep at some time t.[7] In general, however, we can say that no physicalist theory can describe a system which is both omega-capable and capable of reasoning using predicate modal logic, without being inconsistent.

A few minor points should be cleared up here. Since the condition $X \neq R$ will be satisfied even if X does not exist, it is not necessary for M to "believe" the physicalism postulate; in other words, it is not necessary to assume the postulate:

$$\sim\Diamond\sim(W_M(\Psi \supset \Phi) = \text{TRUE})$$

This is an important point, because one of the most common reasons for misunderstanding the proof turns on an implicit assumption of exactly this postulate.[8] A second point is that we do not have to assume that every real process X is a either process of type P or of type R. We only need to assume:

$$(X = R) \supset (X \neq P)$$

not

$$(X = R) \leftrightarrow (X \neq P)$$

One might be tempted to assume that even the first condition might be violated according to processes operating according to "quantum logic", according to which X might be a process of type P *and* a process of type R *simultaneously.* But this only works if one is prepared to allow W(p) to be both TRUE and UNTRUE simultaneously, and such an interpretation of an observable physical state has no meaning in any logic, quantum or otherwise[9].

A final point, which is subtle but extremely important, should also be made; the theorem applies strictly not to systems, but to *theoretical models* of systems. It does not really mean anything to say that no physical system can be omega-capable. All one can say is that no *physicalist model* of a system can be omega-capable. This is important, because many philosophers of mind are deeply unhappy at the idea that such a limitation on the power of physical systems can be proved theoretically. Indeed, they are right to be unhappy about this, but that is not actually what we have demonstrated. What we have shown is that no *physicalist theory* can describe any system which is both omega-capable and capable of reasoning using predicate modal logic without being inconsistent. This subtle but significant difference should never be forgotten or overlooked.

---

[7] The precise Cartesian cogito requires a little more work, since the definition of a type R process for such an omega function is actually self-contradictory -- it clearly makes no sense to define a process which inaccurately tells its owner that it exists. For the cogito one needs to think not in terms of the *accuracy* of processes, but in terms of the *existence of processes.* For an illustration of how this can be done, see the Cartesian lemma from Reason (2019).

[8] This is my experience from a variety of personal communications.

[9] There are paraconsistent logics whose interpretations include states such as "true + false", but in these interpretations, true and false do not have their usual classical meaning.

**Some Misconceptions about the Three Worlds Problem, and what they Reveal about the Limits of Philosophical Analysis**

In what follows, I would like to be able to cite published accounts of how philosophers have responded to the challenge posed by the no-supervenience theorem. Unfortunately I cannot do this, since despite my requests for them to do so, no philosopher has been prepared to submit their response for publication. So the discussion which follows is heavily reliant on personal communications and reviews, most of which are of course unattributable. But certain misconceptions have turned out to be fairly common in those responses I have received. Let us now analyze some of these.

A surprisingly common claim is that the proof relies on an assumption of *infinitism*, that any certain belief must imply an infinite series of justifications. This assumption is not actually made in the proof. The infinite series of physical processes is actually *induced* from a set of much weaker assumptions. These assumptions are the requirement that physical systems must be objective in some way, and the assumption that no physical process can be assumed to be reliable and accurate on an a priori basis. This implies that the accuracy of any physical process must be established empirically, and it is this, together with the assumption of physicalism, which induces the infinite regress. This example shows how important it is to pay attention to the specific inferences and assumptions of the argument, which is best done when the argument is presented mathematically. It is an example of how mathematical argumentation can be far more rigorous than philosophical argumentation.

David Papineau has pointed out that it is not necessary to solve, or even address, the three world problem in order to know that we have Cartesian self-awareness, and that indeed, most people would not even think of doing so (Papineau, personal communication). This is obviously right. But in that case, our ability to be self-aware in a Cartesian sense (or for that matter, any other sense) would be entirely contingent on our *not* addressing or solving the three world problem. As soon as any of us becomes aware of the three world problem, our Cartesian self-awareness would collapse. The three world problem is, in effect, the Godel sentence for any self-aware physical system capable of intelligent reasoning. The logical implication of the three world problem, is that any intelligent system capable of understanding the three world problem can deduce (provably and correctly) that it may either be describable by some physical theory, or self-aware, but never both.

Some philosophers argue that not all physical processes may be objective, and that if there exist physical processes which are not objective, then the proof is not sound. This is a striking example of the difference between reasoning using first-order logic, and reasoning using set theory. In first-order logic the argument can be represented like this:

The objectivity of all physical processes implies that physicalism is false
All physical processes are objective
Therefore physicalism is false

Put like that, it looks as though the entire argument becomes unsound when the second premise is denied. But that is actually not an accurate representation of the argument at all. In set theory, the argument *actually* looks like this:

Let S be the set of all theories in which all physical processes are objective
Any member of S is inconsistent with physicalism
Therefore if Theory T is a member of S, then T is inconsistent with physicalism

There may still be possible physicalist theories in which not all physical processes are objective, but we can still rule out physicalism for every theory in which all physical process *are* objective. This is an example of how arguments expressed in terms of set theory can be far more powerful than arguments expressed in first-order logic. Furthermore, any workable physicalist theory would have to include at least one example of a non-objective physical process which does not supervene on any objective process, and the definition of such an entity is likely to pose a significant challenge to future theoreticians.

Most other comments have assumed that the three worlds argument is defeasible in some way, by postulating specific sub-types of physicalist theory to which the proof is assumed not to apply. This, however, cannot work in any monotonic logic. For example, one correspondent suggests that a viable physicalist theory might depend, not only on the physical processes within an individual brain, but on the physical processes distributed across an entire community of brains. But since the proof applies to *any* set of physical processes, no matter how they are distributed between individuals, the consequences of the proof cannot be evaded in this way. This is another example of how an argument in first-order logic is simply inadequate for dealing with an argument derived from set theory. The same applies to arguments based on evolution, which assume that the relevant physical systems simply emerge through natural selection. It is indeed theoretically possible that such systems might evolve, but if they are capable of intelligent reasoning, they will still be able to deduce that the three worlds problem leads to a contradiction. Once the bomb comes into existence, by whatever means, there is nothing anyone can do to stop it ticking.

A similar problem arises with attempts to redefine or limit the definition of Cartesian self-awareness, so that the proof can be assumed not to apply to some definition of self-awareness preferred by the correspondent: for example, arguments to the effect that the Cartesian Cogito relies on the concept of self or "I". This is yet another example of the confusion between set theory and first-order logic. Cartesian self-awareness, as defined in the three-world problem, refers to the set of all possible interpretations of Cartesian self-awareness, whether they include notions of self or not. There is by definition no way of understanding Cartesian self-awareness which is not a member of the set of all possible interpretations. The contradiction inherent in the three-world problem arises from the syntactic properties of ω (or P), and not on the interpretation of ω (or P). Therefore it makes no difference at all what P is understood to mean. Once again, the metaphysics of self-awareness turns out to be completely irrelevant.

Another surprisingly common response is to argue that, since no-one can get around to wondering about their existence unless they first exist to do the wondering, then the fact that a system is asking questions about its own existence shows that it must exist in the first place. This is actually an example of Slezak's fallacy of the Cartesian spectator (Slezak, 2003). It is obvious to us, the Cartesian spectator, that a system which is pondering its own existence must exist in the first place, but that does *not* mean that such a system is actually aware that it is pondering its own existence. In fact, any belief the system may have about its own pondering will simply generate its own three worlds problem. Any intelligent system will be able to deduce that if it is a physical system, then all of its beliefs about anything must be undecidable. The only way to prevent a physical system from reasoning thus is to limit its ability to reason intelligently about itself.

**Discussion**

It is evident from private discussions I have had with various philosophers and other consciousness researchers, that many of them are deeply reluctant to accept this result. There appear to be three main reasons for this. Firstly, the result rules out a thesis, that there exist viable physicalist theories of consciousness, which decades of philosophical research have been dedicated to defending. Secondly, the above theorem is a definitive result in a field in which, probably, not all that many people expect there to be definitive results of any sort. And thirdly, because the result is a *theorem* about a class of theories and not a theory in itself, it relies on a method of proof which many consciousness researchers are simply not familiar with. Hence the necessity for the present formalizable proof. It should also be noted that a *formalizable* proof is not the same thing as a *formal* proof. A formalizable proof simply presents the inferential steps in sufficient detail that the formal proof of each step should be a mere technical detail; but it is a matter of judgment exactly when this is achieved.

A no-go theorem can be thought of as a proof in some language L, that no theory T expressed in L, can have a certain set of properties. In this case, L is predicate modal logic, and T is any member of the class of all physicalist theories (or, more generally the class of all *supervenient* theories). The proof of such a theorem must be expressed much more rigorously and precisely than is usually the case with philosophical arguments. The proof used here is of a type known in mathematics as a *diagonal method*, for which no satisfactory analog exists in analytic philosophy of mind.

We have stipulated that M must be capable of using predicate modal logic. Although it is not really necessary, it makes the proof clearer and easier to follow if we also allow M to assume that it can rely on its ability to use predicate modal logic. This simply means we can ignore any questions which arise as to the reliability of Wolpert states representing facts which are both a priori and necessary. This may seem somewhat contrived and artificial, but the consequence of not doing this is that *every* proposition then becomes undecidable for M, and hence by default M cannot be omega-capable. So stipulating that M rely on its use of predicate modal logic turns out to be the conservative choice.

*Phenomena.* **237** (9), pp 1257-1281.